\documentclass[prl,twocolumn,superscriptaddress]{revtex4}

\usepackage{amsmath}    
\usepackage{amssymb}    
\usepackage{graphicx}   
\usepackage{verbatim}   
\usepackage{color}      
\usepackage{subfigure}  
\usepackage{hyperref}   
\begin{document}

\title{Conductivity of $\mbox{Si}\mathbf{(111)\operatorname{-}7\times7}$: the role of a single atomic step}
\author{Bruno V.C. Martins}
\affiliation{National Institute for Nanotechnology, National Research Council of Canada, Edmonton, Alberta, Canada}
\affiliation{Department of Physics, University of Alberta, Edmonton, Alberta, Canada T6G 2E1}
\author{Manuel Smeu}
\author{Hong Guo}
\affiliation{Centre for the Physics of Materials and Department of Physics, McGill University, Montreal, Quebec, Canada}
\author{Robert A. Wolkow}
\email{rwolkow@ualberta.ca}
\affiliation{National Institute for Nanotechnology, National Research Council of Canada, Edmonton, Alberta, Canada}
\affiliation{Department of Physics, University of Alberta, Edmonton, Alberta, Canada T6G 2E1}
\email{Corresponding author
rwolkow@ualberta.ca}

\begin{abstract}
The $\mbox{Si}(111)\operatorname{-}7 \times 7$ surface is one of the most interesting semiconductor surfaces because of its complex reconstruction and fascinating electronic properties. While it is known that the $\mbox{Si}\operatorname-7 \times 7$ is a conducting surface, the exact surface conductivity has eluded consensus for decades as measured values differ by 7 orders of magnitude. Here we report a combined STM and transport measurement with ultra-high spatial resolution and minimal interaction with the sample, and quantitatively determine the intrinsic conductivity of the $\mbox{Si}\operatorname{-}7 \times 7$ surface. This is made possible by the capability of measuring transport properties with or without a single atomic step between the measuring probes: we found that even a single step can reduce the surface conductivity by two orders of magnitude. Our first principles quantum transport calculations confirm and lend insight to the experimental observation.
\end{abstract}

\maketitle

The $\mbox{Si}(111)\operatorname{-}7 \times 7$ surface has been extensively investigated for decades because of its complex reconstruction and fascinating electronic properties \cite{takayanagi}. While bulk Si is a semiconductor with an intrinsic band gap, its $7 \times 7$ surface exhibits metallic properties \cite{persson,avouris}. This is due to the adatoms on the surface possessing partially occupied dangling bonds, which form a band that crosses the Fermi level. There have been many experimental attempts to determine the sheet conductivity of this important surface but a consensus has not been reached despite nearly two decades of work. Given the dominating role of Si in micro- and nano-technology, it is indeed surprising that such a basic property as the surface conductivity has evaded consistent measurements for so long. With the continued down-scaling of electronic devices and the advent of nanoelectronics \cite{piva,simmons}, quantitative understanding of surface physics becomes very important and the $\mbox{Si}(111)\operatorname{-}7 \times 7$ surface serves as an excellent testing ground for electron transport at the atomic scale. 

So far the surface conductivity of $\mbox{Si}\operatorname{-}7 \times 7$ has been measured by single tip scanning tunneling microscope (STM), two-point probe (2PP), four-point probe (4PP) and by electron-energy loss spectroscopy (EELS). While EELS reported a value of $3 \times 10^{\operatorname{-}4} \Omega^{\operatorname{-}1} \Box^{\operatorname{-}1}$ \cite{persson}, the STM and multiprobe measurements yielded many different values including $10^{\operatorname{-}4} \Omega^{\operatorname{-}1}\Box^{\operatorname{-}1}$ \cite{yoo}, $10^{\operatorname{-}6} \Omega^{\operatorname{-}1}\Box^{\operatorname{-}1}$ \cite{avouris,hasegawa1}, $4 \times 10^{\operatorname{-}6} \Omega^{\operatorname{-}1}\Box^{\operatorname{-}1}$ \cite{hasegawa3}, $9 \times 10^{\operatorname{-}9}\Omega^{\operatorname{-}1}\Box^{\operatorname{-}1}$ \cite{hashizume,hofmann1}, $9 \times 10^{\operatorname{-}11} \Omega^{\operatorname{-}1}\Box^{\operatorname{-}1}$ \cite{jachinsky}. The dramatic differences between these values could be due to many factors such as the measurement methods being EELS, 2PP or 4PP; the distance between the measuring probes being macro or micro; the doping levels of the Si; the measured nanostructures and quality of samples; as well as the presence or absence of atomic steps in between the measuring probes \cite{janik,hofmann2}. While 4PP is the fundamental technique for transport measurements, it is not capable of achieving high spatial resolution due to the heavy geometrical constraints imposed by the probe size. Therefore, it cannot discern the resistance of a single atomic step, although it does reveal that the surface atomic steps are crucially affecting the measured sheet conductivity \cite{hofmann2,hasegawa4}. In any event, with the experimentally reported values spanning 7 orders of magnitude \cite{hasegawa3,hofmann3}, quantitative understanding of the electronic conduction in $\mbox{Si}\operatorname{-}7 \times 7$ remains an unresolved problem and it is the purpose of this work to fill this gap. 

In particular, we have developed a multiprobe technique for high spatial resolution spectroscopy based on a combination of STM and electronic transport measurements. The resolution is so high that the conductance of a single atomic step can be quantitatively determined and compared with that of the surface having no atomic step at all. This way, we obtain not only the intrinsic sheet conductivity of the $7 \times 7$ surface but also the surprising result that even a single atomic step can drop this conductivity by two orders of magnitude. Since (to the best of our knowledge) this is the first time that resistance of a single atomic step on any semiconductor surface is directly determined by a transport measurement, our results suggest that the dramatic differences in the reported sheet conductivity of the $7 \times 7$ may very well be due to the presence of an unknown number of atomic steps between the measuring points. By carrying out first principles transport simulations we have calculated the effects of a single atomic step on $\mbox{Si}(111)\operatorname{-}7 \times 7$, and the results confirm the observation that a single step has a dramatic effect on the sheet conductivity. Finally, our measured sheet conductivity of the intrinsic (without any step) $\mbox{Si}(111)\operatorname{-}7 \times7$ surface is $(1.3 \pm 0.3) \times 10^{\operatorname{-}6} \Omega^{\operatorname{-}1} \Box^{\operatorname{-}1}$, sitting on the high value side of the reported spectrum of data.
%

We carried out the experiment using our homemade room-temperature multiprobe \cite{multiprobe} instrument which comprises three independent STMs: two side probes (L and R), which contact the surface as current terminals, and a central probe (C) used for imaging and for voltage measurements. The three probe instrument is capable of emulating four-probe measurements by using two fixed probes and sequentially measuring voltages on different surface spots using a third probe referenced to the common ground. The conventional approach for conductivity measurements was not used. Instead a novel method was developed for non-contact voltage measurement using height-bias (ZV) spectroscopy \cite{berndt,zandvliet}. This approach, which was adapted from the set of established STM spectroscopic techniques, consists of monitoring the tip height as a function of the sweeping bias while the control loop is kept closed. At some point during the tip voltage sweep, the sweeping bias on the tip will match the sample potential, causing the tip to get very close to the surface to compensate for the drop in current. It is straightforward to note that if the sample is not grounded and the sample potential is unknown, the ZV spectrum will present a dip corresponding to the surface voltage. It is notable that measurement of the local potential was possible without contact. This is a great advantage since we can move the tip around and construct high spatial resolution and non-destructive maps of the local potential at the surface.

\begin{figure}[ht!]
   \includegraphics[width=8.5cm]{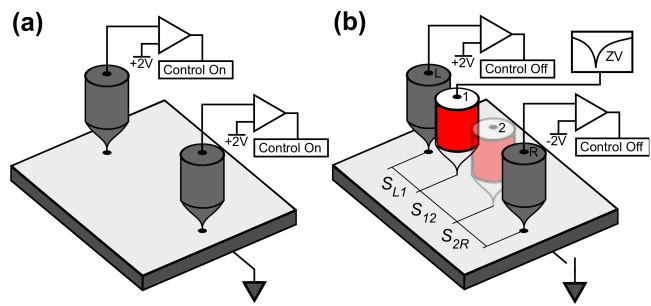}
   \caption{ (color online) (a) Contact control using the STM Proportional-Integral loop with a 2 $\mu$A setpoint. Both probes have bias set to +2V while the sample is grounded. (b) Acquisition of the potential profiles. Control loops of the two side probes (L and R) are turned off, the sample is disconnected from ground and a potential difference of 4 V is established. A current now flows between L and R  and the central probe enters in STM mode for the acquisition of the ZV spectra. After acquisition, the central probe is retracted and the system returns to the condition presented in (a). The potential profile is constructed by repeating this same process at many points along a straight line defined in the STM scan frame. Labels 1 and 2 correspond to the extreme points of the potential profile.}  
   \label{fig:schematic}
\end{figure}

We now describe the experimental approach in detail. Using two side probes labelled L and R to touch the surface and establish a potential gradient, a current flows in between them. It is required that the contact resistance of both probes be the same and constant with time. This was achieved by using STM Proportional-Integral control of each of the probes in order to keep current constant while in contact. As showed in Fig. \ref{fig:schematic}(a), we applied the same control system (one for each probe) used for the STM mode to control the contact. While keeping the sample grounded we applied a common 2V bias to both L and R probes (by using a floating preamp configuration) and established a 2$\mu$A set point for the current. Using this configuration, it was possible to maintain both L/R probes with a constant contact resistance indefinitely. 

Fig. \ref{fig:schematic}(b) shows the steps taken during the measurement process: i) control loops of both L and R probes are turned off and sample is disconnected from ground; ii) a potential gradient is applied between the L, R probes; iii) The C probe enters in the tunneling regime; iv) ZV spectroscopy measures the sample local potential; v) The C probe is retracted and moved to the next measurement position; vi) sample is reconnected to ground and the control loops of both side probes are reconnected. The entire process takes only a few seconds allowing us to disregard contact variations during the measurement.

We applied this technique to measure the conductivity of both a step-free surface and a surface region containing a single monatomic step on $\mbox{Si}\operatorname{-}7 \times 7$. We performed experiments using n-type samples with nominal resistivity of $1 \,\operatorname{-} 3 \, \Omega \operatorname{-} \mathrm{cm}$.

\begin{center}
\begin{figure}[ht!]
   \includegraphics[width=16cm]{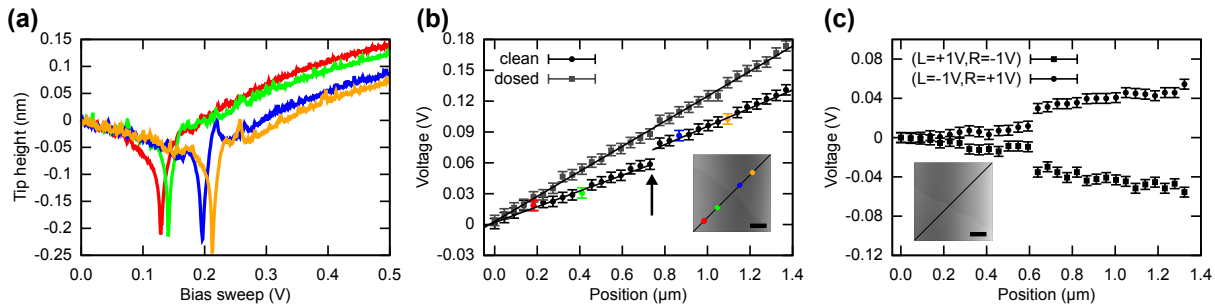}
   \caption{(color online) (a) ZV spectra collected at different positions on the diagonal. (b) potential profile acquired on both clean and dosed (50 L 1,2,4-Trimethylbenzene) surfaces. The clean profile displays the fitting function used to calculate the conductivity. The potential difference is 4V. Inset: STM image showing the diagonal where the profile was acquired and the monatomic step edge. Some points are identified by colours that match the corresponding potential profile displayed in (a). (c) potential profile acquired on clean sample with potential difference 2V and with inverted polarities. Note the mirroring of the profile and the scaling of the voltage by a factor of approximately 2. All scale bars = 200nm.}
   \label{fig:profiles}
\end{figure}
\end{center}

By fixing the L/R probes on the sample surface with or without a single atomic step in between, and applying a total bias voltage of 4V ($\mbox{L}  =  \operatorname{-}2 \mbox{V} , \mbox{R}  =  +2 \mbox{V}$), we record the ZV spectra by the central probe C at multiple points along the surface (see Fig. \ref{fig:schematic}(b)). Fig. \ref{fig:profiles}(a) shows four recorded typical ZV curves versus the potential of probe C for a surface with a single atomic step, each curve at a unique surface position as indicated by the colored symbols in Fig. \ref{fig:profiles}(b). Each point in Fig. \ref{fig:profiles}(b),was calculated from an average of 10 ZV spectra. The most prominent feature of the ZV curves is the very sharp dip, the position of which gives a measure of the local sample potential as discussed above. In this way the potential value was collected at many positions on the surface. Measurements traversing a single atomic step are plotted in Fig. \ref{fig:profiles}(b). A most important finding is the step-like jump (indicated by the up-arrow) in the voltage profile which correlated well with the STM step topography (inset). We observe that, due to surface conduction from L to R, the resistance of a single atomic step creates a potential jump which our experiment measures unambiguously. 

To further confirm that the potential jump was indeed due to surface conduction, we acquired another potentiometric curve (see Fig. \ref{fig:profiles}(b): this time on a sample dosed for 50 s at $10^{-6}$ Torr with 1,2,4-Trimethylbenzene. By STM imaging, we found that such a dose reacts with most of the surface dangling bonds, thus reduces (or eliminates) the surface conduction mechanism. Indeed, we observe an increase in slope compared to that for the clean (un-dosed) sample and the potential jump has disappeared even though the atomic step remains visible in the STM topography. The disappearance of the potential jump indicates the loss of a surface-state conduction channel. The overall increase in slope after dosing the silicon dangling bonds corresponds to a net decrease of conduction. After dosing, only the relatively resistive space charge layer and bulk Si remain available for conduction. 

By reversing the polarity of the bias voltage applied to the L/R probes, the potential drop also reverses as shown in Fig. \ref{fig:profiles}(c). Such an inversion of the voltage jump and the slope confirm that the observed physical feature (e.g. the potential jump) is not topographical in origin but from conduction on the surface. Measurements were also carried out on a highly-dopde sample (see the Supplemental Material \cite{supp} for further details).

\onecolumngrid

\twocolumngrid
Having established that the potential jump is due to the single atomic step in the path of surface conduction between the L/R probes, we now determine the surface conductivity. From Fig. \ref{fig:profiles}(b), we fitted the potential profiles for the un-dosed case, before and after the step separately. By extending these functions along the length of the diagonal, we calculated two potential drops which are averaged, giving us a total voltage drop without the step of $\Delta \mathrm{V} = (0.110 \pm 0.005) \mathrm{V}$.  The total current between the L/R probes was measured to be $I = (4.57 \pm 0.03) \mu A$. Equivalently, we fitted the potential profile to the dosed case and obtained a voltage drop of $\Delta \mathrm{V} = (0.170 \pm 0.005) \mathrm{V}$ and total current $I = (3.25 \pm 0.03) \mu A$. For four-probe surface conduction, the sheet conductivity with variable spacing is given by the following formula \cite{hofmann2}:
\begin{equation} \centering
   \sigma = \dfrac{I}{2 \pi \Delta V} \left( \ln \dfrac{S_{12}+S_{2R}}{S_{L1}} + \ln \dfrac{S_{L1}+S_{12}}{S_{2R}} \right)
   \label{eqn:surface} 
\end{equation}
where $S_{\alpha \beta}$ are distance parameters defined in Fig. \ref{fig:schematic}(b) and for our experiments, $S_{L1}=(5 \pm 1) \mu m$, $S_{12}=(1.00 \pm 0.01) \mu m$ and $S_{2R}=(6 \pm 1) \mu m$. We obtain $\sigma_c=(2.7 \pm 0.4)\times 10^{\operatorname{-}6}\Omega^{\operatorname{-}1}\Box^{\operatorname{-}1}$ from Eq. \ref{eqn:surface} for the surface without any atomic step in between the L/R probes. This measured conductivity value is a combination of surface, space charge layer and bulk conductivities. Incidentally, this value is consistent with those sitting at the higher end of the measured Si-$7 \times 7$ sheet conductivity spectrum as discussed above \cite{avouris,hasegawa1,hasegawa3}.

To extract a surface conductivity we consider the system before and after dosing. After dosing, the surface state contribution was eliminated and only space charge layer and bulk remained. The dosed surface conductivity is $\sigma_d=(1.4 \pm 0.2)\times 10^{\operatorname{-}6}\Omega^{\operatorname{-}1}\Box^{\operatorname{-}1}$. The difference yields the surface conductivity of $\sigma_s=(1.3 \pm 0.3)\times 10^{\operatorname{-}6}\Omega^{\operatorname{-}1}\Box^{\operatorname{-}1}$. Note that, using this method, we were able to determine the surface conductivity despite of the complexities involved in the analysis of bulk conduction in semiconductors \cite{hofmann4,hasegawa5,wells}.

Next, we analyze the tantalizing question: what is the conductivity of a single atomic step? From the un-dosed data of Fig. \ref{fig:profiles}(b), the potential jump (pointed by up-arrow) is $\Delta \mathrm{V}_S = (0.018 \pm 0.005) \mathrm{V}$, and this value is used in Eq.\ref{eqn:surface} for the parameter $\Delta \mathrm{V}$. From the difference between dosed and un-dosed currents, we obtained the surface current $I = (1.32 \pm 0.03) \mu A$ To fix the distance parameter $S_{12}$ (see Fig. \ref{fig:schematic}(b)), we note Ref. \cite{hupalo1} showed altered electronic structure in the vicinity of a step on the $7 \times 7$ surface in a region approximately 1.5 nm wide. It is thus reasonable to set $S_{12} = 1.5 \mathrm{nm}$ since the potential jump at the atomic step should occur across such a distance. With these values and by Eq. \ref{eqn:surface},  we obtain the atomic step conductivity $\sigma_{step} = (5 \pm 3) \times 10^{\operatorname{-}8} \Omega^{\operatorname{-}1}\Box^{\operatorname{-}1}$. This value is two orders of magnitude smaller than the measured surface conductivity without any atomic step, and indicates that the surface conductivity of the $\mbox{Si}\operatorname{-}7 \times 7$ can drop by almost a hundred times even if there is only a single atomic step sitting in between the L/R probes. We conclude that the conducting properties of the $\mbox{Si}\operatorname{-}7 \times 7$ surface is very sensitively influenced by the individual low conductivity contributions of the atomic steps.

\begin{figure}[ht!] \centering
   \includegraphics[width=8cm]{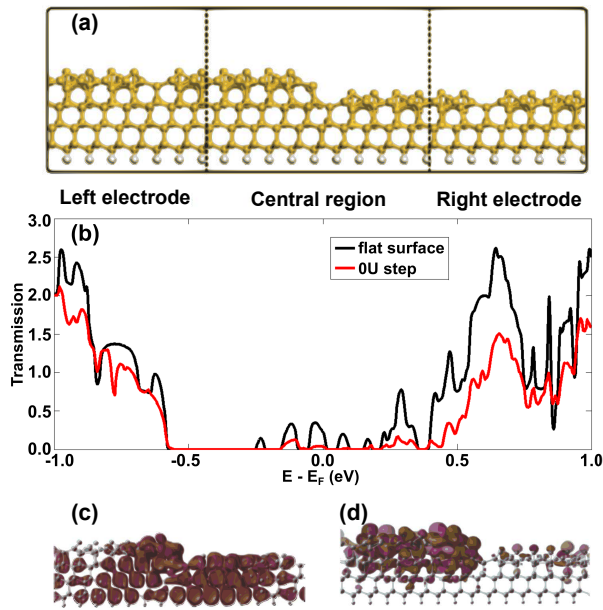}
   \caption{(color online) (a) Two-probe structure used to calculate electron transport through a step on $\mbox{Si}\operatorname{-}7 \times 7$. Flat $\mbox{Si}\operatorname{-}7 \times 7$ surface makes up the electrodes. (b) Transmission spectra of a flat surface and one with an atomic step. Bottom: scattering states through $0U$ step. (c) is for the transmission at $-0.6$ eV, (d) is for the small peak near $E_F$ at -0.03 eV in (b).}
   \label{fig:theory}
\end{figure}

To confirm this understanding, we have carried out first principles calculations of electron transport along the $\mbox{Si} \operatorname{-}7\times7$ surface with and without a single atomic step. Due to the complexity of this surface there are 14 distinct types of atomic steps possible, depending on whether the edge of the top terrace (of the step) corresponds to the faulted or unfaulted half of the $7\times7$ supercell and the overlap of the upper and lower terraces, as was described by Tochihara et al.\cite{tochihara}. Here we opted to focus on the 0U step type \cite{hupalo2}. The atomic positions were relaxed by density functional theory (DFT). The obtained structural and electronic properties are consistent with Ref. \cite{hupalo2}. More details can be found in the Supplemental material \cite{supp}. The relaxed atomic step structure was then used to construct two-probe transport junctions shown in Fig. \ref{fig:theory}(a) for electron transport calculations based on the non-equilibrium Green's function technique (NEGF) combined with DFT. The NEGF-DFT method is a well-established formalism for first principles analysis of charge transport, for details we refer interested readers to Refs \cite{guo1,guo2,smeu1}.

Fig. \ref{fig:theory}(b) compares the transmission spectra for a surface having a single $0U$-step and another one having no step which is a flat $7\times7$ surface having 6 Si layers. The conductance of the system is given by $G=TG_0$ where $T$ is the transmission coefficient and $G_0=2e^2/h$ is the conductance quantum. As expected, the transmission through the flat surface without any atomic step (black solid curve) has a significant value near $E_F$ (set at energy zero) which is attributed to the electronic states originating from adatoms and possibly dimer atoms \cite{smeu2} of the $7\times7$ supercell. These are surface states due to the $7\times7$ reconstruction and they lie inside the Si band gap bounded by the valence band at -0.6 eV and the conduction band near 0.4 eV. When an atomic step is present on the surface, the calculated transmission inside the band gap is greatly reduced (red curve), this behavior agrees with our experimental observation. 

The surface conduction is vividly shown by plotting the calculated wave functions of the scattering states in real space. Scattering states are eigenstates of the open device Hamiltonian connecting the two measuring probes through the scattering region (see Fig. \ref{fig:theory}(a).  Two such scattering states (at two different energies) are plotted Fig. \ref{fig:theory}(c),(d). Fig. \ref{fig:theory}(c) is a scattering state at energy $-0.6 \mbox{eV}$ which is via the valence band of bulk Si. This wave function is spread throughout the $7\times7$ slab indicating a large transmission, consistent to that of Fig. \ref{fig:theory}(b). Fig. \ref{fig:theory}(d) is for a scattering state at the energy just below the $E_F$ and this state does not transmit well to the right electrode indicating a much smaller transmission, also in agreement with Fig. \ref{fig:theory}(b). Importantly, the scattering state in Fig. \ref{fig:theory}(d) is predominantly located on the top of the surface and does not penetrate into the bulk layers, confirming the surface conduction in the bulk band gap. Fig. \ref{fig:theory}(d)  gives a clear picture of the resistive nature of the atomic step: the amplitude being much smaller for electrons having passed through the step (right side of Fig. \ref{fig:theory}(d)). Finally, we have carried out calculations on surfaces with other types of atomic step ($0F$, $6F$, $2U$). While these atomic steps give slightly different transmission spectra near $E_F$, every type reduces the transmission at $E_F$ by $3 \operatorname{-}6$ times than that through the flat $7\times7$ surface without any step. Our first principles modelling thus reaches the same conclusion as our experimental measurements, namely the presence of a single atomic step drastically reduces the overall surface conductance of Si-$7\times7$.

In summary, we have developed a combined STM and transport measurement technique with ultra-high spatial resolution and minimal interaction with the sample, to directly and quantitatively determine the intrinsic conductivity of the $\mbox{Si}(111)\operatorname{-}7\times7$ surface to be $(1.3 \pm 0.3)\times 10^{\operatorname{-}6}\Omega^{\operatorname{-}1}\Box^{\operatorname{-}1}$. This is made possible by the capability of measuring transport property with or without a single atomic step between the measuring probes. In particular, we found that even a single atomic step can drop the surface conductivity by two orders of magnitude. Our first principles quantum transport calculations confirm the experimental observations. Our reported experimental technique overcomes a serious limitation of conventional four-probe method, namely it has not been possible to get the four-probes close enough to resolve the contribution to conduction by a single atomic step. The new technique is very well-suited for surface spectroscopy studies of nano-systems and opens the door for direct investigations of charge transport in a vast array of nanostructures. 

We thank Prof. Wei Ji for his participation in the early stages of this work and CLUMEQ and RQCHP for access to computational resources. We are grateful for the able technical assistance of Martin Cloutier and Mark Salomons. This work was supported by NRC, NSERC and iCORE of Canada.	

\end{document}


\noindent EPAPS: Supplemental Material for ''Conductivity of $\mbox{Si}(111)\operatorname{-}7 \times 7$: the role of a single atomic step'', by Bruno V.C. Martins, Manuel Smeu, Hong Guo, and Robert A. Wolkow
\\
\\
\noindent In this Supplemental Material, we present technical details of the experimental measurements and theoretical calculations. 

\section{Experimental measurements}
In our measurements, we used n-type samples with nominal resistivity $1\operatorname{-}3 \, \Omega\operatorname{-}\mathrm{cm}$ (Low-doped - LD) and $0.001\operatorname{-}0.003 \, \Omega \operatorname{-} \mathrm{cm}$ (Highly-doped - HD). Both sample surfaces were prepared in UHV using a sequence of $1250^o \mathrm{C}$ fast heating cycles. The $7 \times 7$ reconstruction was obtained by adding one extra 1 minute step at $850^o \mathrm{C}$ to the last cycle followed by a slow cool down (2 min) to room temperature.

\section{Measurements with Highly-doped samples}

\begin{center}
\begin{figure}[ht!]
   \includegraphics[width=6cm]{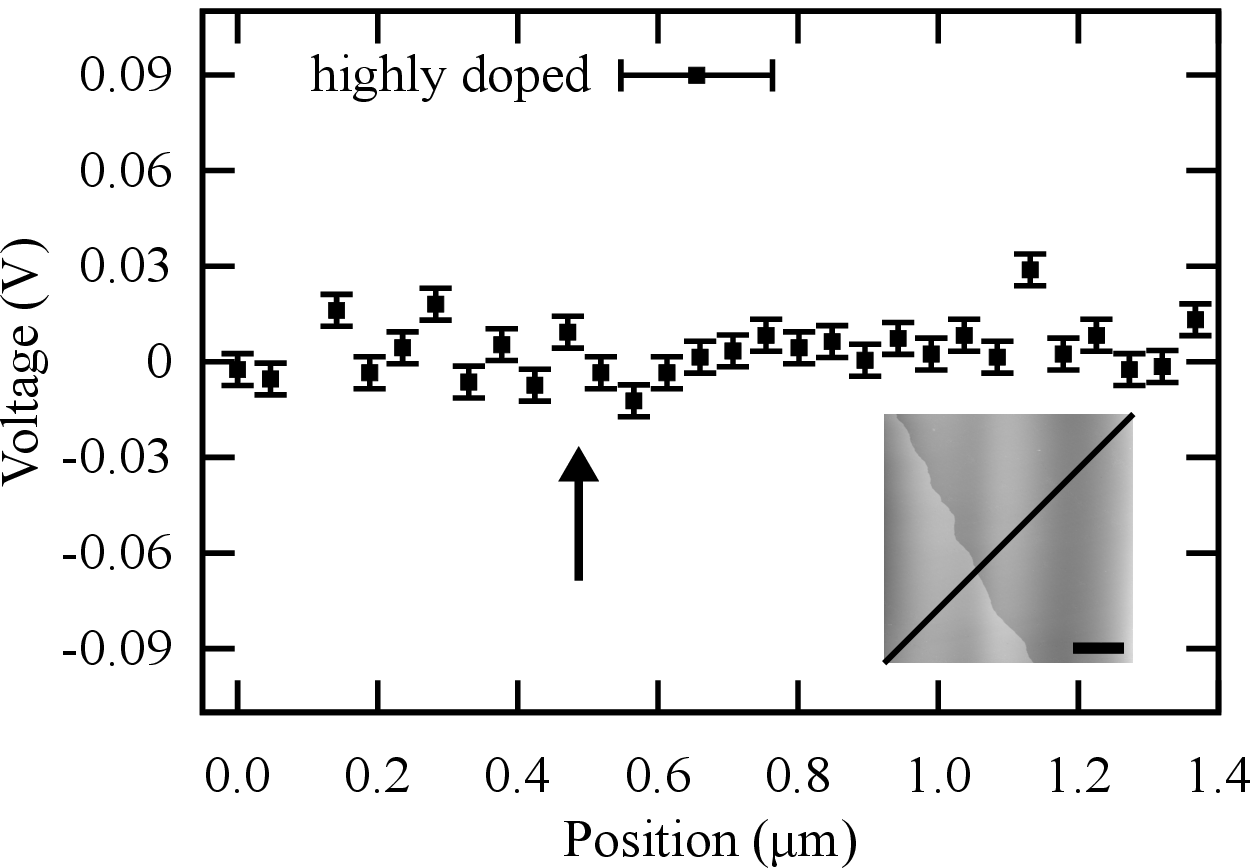}
   \caption{Highly-doped potential profile. The arrow marks the position of the atomic step, which is visible in the STM topographic image but not in the potential profile. Scale bar is 200nm.}
   \label{fig:hd_stm} 
\end{figure}
\end{center}

A potential profile on an HD sample, as shown in Fig. \ref{fig:hd_stm}. Because the bulk conductivity is relatively high, our voltage probe spacing of $\approx 1 \, \mu \mbox{m}$ is insufficient to reveal a voltage drop. Not surprisingly then, no voltage drop at a single step can be seen either. The barrier which effectively isolates surface states from the bulk on a low doped sample is sufficiently thin for highly-doped samples so as to permit tunneling between surface and bulk.


\section{Theoretical calculations}
\begin{figure}[ht!] \centering
   \includegraphics[width=6cm]{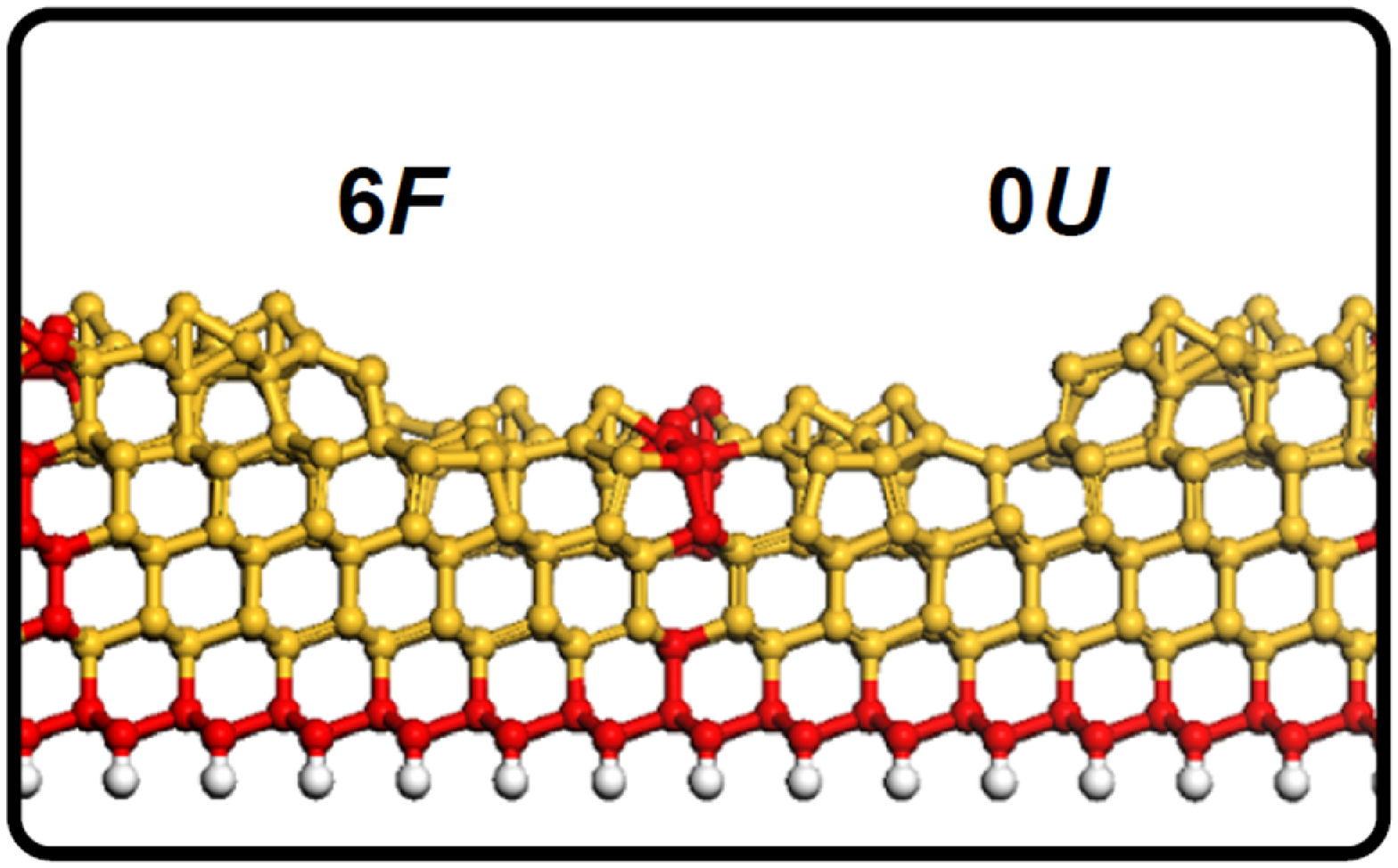}
   \caption{Periodic structure used for relaxation of atomic step on $\mbox{Si}\operatorname{-}7\times7$. Si atoms shown in red were frozen to their relaxed positions in a slab with no step, the bottom bilayer is frozen to bulk Si positions.}
   \label{fig:theory}
\end{figure}

In our theoretical calculations, the system shown in Fig. \ref{fig:theory} was constructed for structural relaxation. The presence of two steps was necessary for periodic boundary conditions of the density functional theory (DFT) calculation. The relaxations were carried out with the Vienna {\it ab initio} simulation package (VASP) \cite{kresse1,kresse2} using the Perdew-Burke-Ernzerhof generalized gradient approximation (PBE-GGA) for the exchange correlation energy \cite{perdew}. A projector augmented wave method was used for the ionic potentials \cite{blochl,kresse3} with a kintetic energy cutoff for the plane wave basis of 400 eV. The atoms shown in red along the bottom of the slab were frozen to their bulk Si positions and the other red atoms were frozen to their respective positions for a flat $\mbox{Si}\operatorname{-}7\times7$ surface (with no step). Note that the dangling bonds at the bottom of the slab were saturated by H atoms. The structure relaxation was deemed complete when the net force on each atom (except the frozen red atoms in Fig. \ref{fig:theory}) was less than $0.02 \, \mbox{eV}/\mbox{\AA}$.  A minimum of $9 \, \mbox{\AA}$ of vacuum space separated the atoms of one supercell from atoms of its image (next repeat unit) in the direction normal to the surface. In the plane of the surface, the Brillouin zone was sampled with sufficient $k$ points so that the energy was converged to less than 1 meV/atom.

During the NEGF-DFT self-consistent calculations of quantum transport, the retarded Green's function at energy $E$ is obtained by the following formula, $G(E)=\left[(E+i \eta)S-H-\Sigma_L-\Sigma_R\right]^{-1}$, where $H$ and $S$ are the Hamiltonian and overlap matrices for the central region of the two-probe device structure (see Fig. 3a of the main text), calculated by DFT. $\Sigma_{L,R}$ are self-energies that account for the effect of the left and right electrodes on the central region and $\eta$ is a positive infinitesimal. The self-energy is a complex quantity with its real part representing a shift of the energy levels and its imaginary part representing their broadening, which can be expressed as the line width matrix, $ \Gamma_{L,R}=i\left(\Sigma_{L,R}-\Sigma_{L,R}^\dagger\right) $.  The self-energy is calculated within the NEGF-DFT formalism by an iterative technique \cite{sancho}. From these quantities, the electronic density matrix can be obtained as,
\begin{equation}
\rho = \dfrac{1}{2\pi}\int_{-\infty}^{\infty} \left[ f(E,\mu_L) G \, \Gamma_L \, G^{\dagger} + f(E, \mu_R) G \, \Gamma_R \, G^{\dagger} \right] dE
\end{equation}
where $\mu_{L,R}$ are the electrochemical potentials of the left and right electrodes and the Fermi-Dirac function that describes the population for a given energy and electrochemical potential. The density obtained from the above equation is used in a subsequent NEGF-DFT iteration step and the cycle is repeated until self-consistency is achieved.  The transmission function is then obtained from the Green's function $T(E) = Tr \left( \Gamma_L G \Gamma_R G^{\dagger} \right)$, which represents the probability that an electron with a given energy $E$ transmits from one electrode through the central region into the other electrode.

In the NEGF-DFT calculations, norm-conserving pseudopotentials \cite{martins} were used to describe the atomic cores, single-$\zeta$ polarized numerical orbitals were used to describe the valence electronic basis, and the exchange-correlation was treated at the local density approximation level. For real space grids, at least 3 grid points for each a.u. were used. The $k$-sampling was carried out on a $3 \times 3$ grid in the plane of the slab for the electrodes, and 3 $k$-points were used for the central region in the direction perpendicular to the page.